\documentclass{aa}

\usepackage{graphicx}

\begin{document}
\title{Simultaneous determination of $\Omega_{\rm M0}$ and $H_0$ from joint
Sunyaev-Zeldovich effect and X-ray observations with median statistics}

\author{Mauro Sereno \inst{1}}

\offprints{M. Sereno, \\
\email{sereno@na.infn.it}}

\institute{Dipartimento di Scienze Fisiche, Universit\`{a} degli Studi di Napoli
``Federico II", Via Cinthia, Compl. Univ. di Monte S. Angelo, 80126 Napoli, Italia}

\date{}

\titlerunning{$\Omega_{\rm M0}$ and $H_0$ from SZE and X-ray observations}
\authorrunning{M. Sereno}

\abstract{
Clusters of galaxies contain a fair sample of the universal baryonic mass fraction. A
combined analysis of the intracluster medium (ICM) within their hydrostatic regions, as
derived from both Sunyaev-Zeldovich effect (SZE) measurements and X-ray images, makes it
possible to constrain the cosmological parameters. We consider both gas fraction estimates
and angular diameter distance measurements. Adopting median statistics, we find, at the
2-$\sigma$ level, the pressureless matter density, $\Omega_{\rm M0}$, to be between 0.30 and
0.40 and the Hubble constant, $H_0$, between 44 and 66~Km s$^{-1}$~Mpc$^{-1}$.
}

\maketitle

\section{Introduction}
Rich clusters of galaxies are the largest known virialized structures in the Universe. Their
importance in observational cosmology is known from early times. In the beginning of the last
century,  direct estimates of their total masses first stated the need for unseen dark matter
(Zwicky~\cite{zwi33}). Now, they continue to provide important information to characterize
the Universe (see, for example, Wang et al.~\cite{wan+al00}; Sereno~\cite{io02cl}).

A method, independent of the nature of the dark matter and based on minimal assumptions, to
constrain the geometry of the Universe and its matter and energy content is based on gas mass
observations in clusters of galaxies.

As shown with X-ray and optical imaging, the main contribution to the baryonic budget in
clusters of galaxies is provided by the hot gaseous intracluster medium (ICM). The gas mass
is about an order of magnitude larger than the mass in stars in cluster galaxies and provides
a reasonable estimate of the cluster's baryonic mass (Fukugita et al.~\cite{fuk+al98}; Lin et
al.~\cite{lin+al03}).

The ICM is X-ray emitting via thermal bremsstrahlung and produces a spectral distortion of
the cosmic microwave background radiation (CMBR), known as Sunyaev-Zeldovich effect (SZE)
(Sunyaev \& Zeldovich \cite{sun+zel70}; Birkinshaw~\cite{birk99}). The SZE is proportional to
the pressure integrated along the line of sight, i.e. to the first power of the gas density.
X-ray emission depends on the second power of the density. The ICM mass fraction may be
estimated from either of these, fundamentally different, observables, based on independent
techniques.

Since there is no efficient way to change the baryon fraction averaged within a Mpc scale,
the mass composition in clusters, when estimated out to a standard hydrostatic radius, is
expected to reflect the universal mass composition (White et al.~\cite{whi+al93};
Sasaki~\cite{sas96}; Evrard~\cite{evr97}). Precise measurements of the visible baryonic mass
fraction in galaxy clusters, $f_{\rm B}$, along with our knowledge of the universal baryonic
density parameter, $\Omega_{\rm B0}$, provides a physically based technique for estimating
cosmological parameters (White et al.~\cite{whi+al93}). Ratio of the ICM mass to the total
mass of the cluster represents a first estimate of $f_{\rm B}$. So, a useful constraint on
the cosmological parameters is given by the identity $\Omega_{\rm M0} =
\Omega_{\rm B0} /f_{\rm B}$. Independent estimates of $f_{\rm B}$ can
be obtained either from SZE measurements or X-ray imaging
observations. The angular diameter distance, where the dependence on
the cosmological parameters appears, to the cluster enters in the two
gas mass estimates through a characteristic length-scale of the
cluster along the line of sight. From the different dependencies of
SZE and X-ray emission on the density of ICM, different dependencies
on the assumed cosmology are implied.

Galaxy cluster gas mass fractions from SZE measurements have been used in Grego et
al.~(\cite{gre+al01}) to constrain $\Omega_{\rm M0}$. X-ray gas mass fraction in relaxed
clusters have provided upper limit on the cosmological density parameter (Mohr et
al.~\cite{moh+al99}; Ettori \& Fabian~\cite{ett+fab99}; Allen et al. \cite{all+al02}; Erdogdu
et al.~\cite{erd+al02}; Ettori et al.~\cite{ett+al02}).

Without referring to the universal baryonic fraction, with some assumptions about the
geometry of the cluster, a joint analysis of SZE measurements with X-ray imaging
observations, since the different density dependencies, makes it possible to determine the
distance to the cluster. Such a distance, independent of the extragalactic distance ladder,
is then used to measure the Hubble constant (Birkinshaw~\cite{birk99};  Mason et
al.~\cite{mas+al01}; Jones et al.~\cite{jon+al02}; Reese et al.~\cite{ree+al02}).

All of the methods discussed above are based on ICM mass observations.
Estimates of gas mass depend on the cosmological parameters through
the angular diameter distance to the cluster. Equating the gas
fraction in cluster to the universal baryonic fraction allows to
investigate both the Hubble constant and the cosmological density
parameters. Taking advantage of the different density dependencies,
SZE and X-ray observations provide independent constraints in the
space of cosmological parameters, leading one to solve for two
unknowns at the same time.

Distance measurements are obtained by equating the central densities
as derived from SZE and X-ray methods. Instead of equating the gas
fraction to an universal value, now the two gas mass estimates are
forced to coincide each other. This constraint is independent and
nearly orthogonal to the previous ones and allows to solve for an
additional unknown.

Usually, when deriving cosmological constraints from gas mass fractions, in order to estimate
$\Omega_{\rm M0}$, a prior on the Hubble constant is assumed. On the other hand, using the
cosmic distance scale from interferometric measurements of the SZE to determine $H_0$
requires, with the current data sets, to fix the background cosmology. We show how a joint
analysis of all the information on the ICM from both SZE and X-ray measurements enables to
determine, at the same time, $\Omega_{\rm M0}$ and $H_0$.

The quality of the data samples used in the analysis determines the area of the overlapping
region and the precision of the estimate. We perform a combined analysis of data samples,
available in literature, of SZE measurements and X-ray observations.

Since we have to combine different data sets, obtained with
independent methods, we adopt median statistics. As shown in Gott et
al.~(\cite{got+al01}, see also Avelino et al.~\cite{ave+al02} and Chen
\& Ratra~\cite{ch+ra03}), median statistics provide a powerful
alternative to $\chi^2$ likelihood methods with fewer assumptions
about the data. Statistical errors are not required to be known and
Gaussianly distributed. Since errors, as reported in X-ray and SZE
literature, are usually asymmetric, performing an analysis without
using the errors themselves turns out to be a very conservative
approach. Furthermore, median statistics is also less vulnerable to
the presence of bad data and outliers. Since we are proposing a method
to constrain cosmology with ICM mass estimates which extend and
combine methods previously established, in this early stage we think
that median statistics can be the best choice.

In Section~\ref{sec2}, we derive the dependence of the ICM gas mass, derived from X-ray
observations or SZE measurements, on the assumed cosmological parameters. In Section 3, the
data samples and our selection criteria on the clusters are presented.  In Section 4, the
median statistics is shortly presented and the results are listed. Section 5 is devoted to
the discussion of various systematic uncertainties. Conclusions are in Section 6.

\section{The gas mass estimate}
\label{sec2}

The ICM mass is calculated by integrating the ICM density profile, $\rho_{\rm ICM}$, over an
assumed shape,
\begin{equation}
\label{frac1}
M_{\rm ICM}(V)=\int_V \rho_{\rm ICM}(\vec{r})d^3\vec{r} \propto m_{\rm
p}\int_V n_{\rm e}(\vec{r})d^3\vec{r},
\end{equation}
where $n_{\rm e}$ is the electron number density profile and $m_p$ is
the proton mass. For a spherically symmetric system, like the widely
used $\beta$-model (Cavaliere \& Fusco \cite{cav+fus76,cav+fus78}),
the mass within a fixed metric radius $r$ is
\begin{equation}
\label{frac2}
M_{\rm ICM} \propto n_{\rm e0}r^3,
\end{equation}
where $n_{\rm e0}$ is the central number density. The radius $r$ can be expressed as the
product of an observable angular radius by the angular diameter distance to the cluster,
$d_{\rm A}$.

The central number density can be measured by fitting the X-ray surface brightness profile,
$S_{\rm X} \propto
\int n_{\rm e}^2
\Lambda (T_{\rm e})dl$, where the integration is along the line of sight and
$\Lambda (T_{\rm e})$ is the X-ray emissivity at the electronic temperature $T_{\rm e}$. The
dependence on distance is made explicit by expressing the integration variable in a
dimensionless form, $dl
= d_{\rm A} d \zeta$. It comes out
\begin{equation}
\label{frac3}
n_{\rm e0}^X \propto \sqrt{\frac{S_{\rm X}}{d_{\rm A}}},
\end{equation}
so, for the gas mass, when measured with X-ray observations, it is
\begin{equation}
\label{frac34}
M_{\rm ICM}^X\propto d_{\rm A}^{5/2}.
\end{equation}

Spatially resolved measurements of the SZE  can also determine $n_{\rm e0}$. Since the
brightness temperature decrement of the CMBR towards a cluster is expressed as $\frac{\Delta
T_{\rm SZ}}{T_{\rm CMBR}}\propto
\int n_{\rm e} T_{\rm e} dl$, then
\begin{equation}
\label{frac4}
n_{\rm e0}^{\rm SZ} \propto \frac{\Delta T_{\rm SZ}}{T_{\rm CMBR}}\frac{1}{d_{\rm A}};
\end{equation}
so, it is
\begin{equation}
\label{frac4b}
M_{\rm ICM}^{\rm SZ}\propto d_{\rm A}^{2}.
\end{equation}
The temperatures generally taken in analyses are the X-ray emission weighted temperatures,
usually measured over a few core radii. If the temperature has a spatial structure, the
temperature inferred from such an average procedure may depend on how much of the cluster is
considered. This can lead to a substantial change in the SZE inferred, and thus to a
systematic error in the determination of the cosmological parameters (Majumdar \& Nath
\cite{ma+na00}).

Typically, an estimate of the total mass of a cluster of galaxies is obtained under the
assumption that the gas, supported solely by thermal pressure, is in hydrostatic equilibrium
in the cluster's gravitational potential. Assuming spherical symmetry and isothermal gas, the
total mass of a cluster within radius $r$ is
\begin{equation}
\label{frac5}
M_{\rm TOT}(<r)=\frac{k_{\rm B} T_{\rm X} r}{G \mu m_{\rm p}}\frac{d
\log n_{\rm e}(r)}{d
\log r}\propto d_{\rm A},
\end{equation}
where $k_{\rm B}$ is the Boltzmann's constant, $\mu m_{\rm p}$ is the
mean molecular weight of the gas and $T_{\rm X}$ is the spatially
averaged X-ray emission temperature of the gas determined from a
broad-beam spectroscopic instrument. Combining Eqs.~(\ref{frac34},
\ref{frac4b}) with Eq.~(\ref{frac5}), the two estimates of the ICM
mass fraction turn out
\begin{equation}
\label{frac6}
f_{\rm ICM} \equiv \frac{M_{\rm ICM}}{M_{\rm TOT}} \propto d_{\rm
A}^{\beta},
\end{equation}
where, for X-ray observations, $\beta_{\rm X}=3/2$ and, for SZE data, $\beta_{\rm SZ}=1$.

By eliminating $n_{\rm e0}$ from Eqs.~(\ref{frac3},~\ref{frac4}), one can solve for the
angular diameter distance, yielding
\begin{equation}
\label{frac7}
d_{\rm A} \propto \frac{\Delta T_{\rm SZ}^2}{S_{\rm X}}.
\end{equation}

The derived values of $f_{\rm ICM}$ depend on the cosmological parameters through the angular
diameter distances (Sasaki~\cite{sas96}). In a Friedmann-Lema\^{\i}tre-Robertson-Walker universe
filled in with pressureless matter and a cosmological constant, the angular diameter distance
to a source at redshift $z$ is
\begin{eqnarray}
\label{dist1}
\lefteqn{d_{\rm A}(z)=\frac{c}{H_0}\frac{1}{|\Omega_{\rm K0}|^{1/2}(1+z)}} \\
& & {\times} {\rm Sinn}
\left\{
\int_{0}^z
\frac{|\Omega_{\rm K0}|^{1/2}}{\sqrt{\Omega_{\rm M0} (1+z^{'})^3+\Omega_{\Lambda 0}+\Omega_{\rm K0} (1+z^{'})^2}}
dz^{'} \right\}, \nonumber
\end{eqnarray}
where $\Omega_{\Lambda 0}$ is the reduced cosmological constant, $\Omega_{\rm K0}=
1-\Omega_{\rm M0} -\Omega_{\Lambda 0}$, and ${\rm Sinn}(x)$ is $\sinh (x)$, $x$, $\sin (x)$
for $\Omega_{\rm K0}$ greater than, equal to and less than zero, respectively; $H_0$ is the
today Hubble constant. For the expression of the distance in inhomogeneous universes we refer
to Sereno et al. (\cite{ser+al01},~\cite{ser+al02}). In the next, we will consider only the
flat case, $\Omega_{\rm K0}=0$, strongly supported by the bulk of evidences (de Bernardis et
al.~\cite{deb+al00}; Harun-or-Rashid \& Roos~\cite{ha+ro01}).

To compare the ICM mass fraction of different clusters, we have to study the same portion of
the virial region in each cluster. In such a region, we expect all clusters to have the same
gas fraction and the gas to be isothermal. Regions of different clusters are characterized by
the same properties if they encompass the same mean interior density contrast, $\delta_{\rm
c}$, with respect to the critical density at their own redshit, $\rho_{\rm c}(z) \equiv
\frac{3H^2(z)}{8 \pi G}$, as strongly suggested by numerical simulation (Evrard et
al.~\cite{evr+al96}; Frenk et al.~\cite{fre+al99}); $H(z)$ is the redshift dependent Hubble
constant,
\begin{equation}
H(z)=H_0 \sqrt{\Omega_{\rm M0} (1+z)^3+\Omega_{\Lambda 0}+\Omega_{\rm K0} (1+z)^2}\ .
\end{equation}
The density contrast $\delta_{\rm c}$ is attained at the radius $r_{\delta_{\rm c}}$ (Evrard
et al.~\cite{evr+al96}; Evrard~\cite{evr97}). According to an analysis of gas velocity
moments (Evrard et al.~\cite{evr+al96}), $\delta_{\rm c} = 500$ is a conservative estimate of
the boundary between the inner, nearly hydrostatic and virialized central region of the
cluster and the surrounding, recently accreting outer envelope. Clusters of different
temperatures have similar structures once scaled to $r_{500}$, within which the easily
visible region of the cluster is also probed with the typical background sensitivity.
Hydrodynamical simulations (Evrard et al.~\cite{evr+al96}) also show that the assumption of
an isothermal gas in hydrostatic equilibrium is valid within the virial radius.

The virial equilibrium expectations at a fixed density contrast, that is
\begin{equation}
\label{vir1}
T \sim \frac{G M}{r_{\delta_{\rm c}}} \propto \frac{\delta_{\rm c}
\rho_{\rm c} r_{\delta_{\rm c}}^3}{r_{\delta_{\rm c}}} \propto
\delta_{\rm c} H^2(z) r_{\delta_{\rm c}}^2,
\end{equation}
can be calibrated by means of numerical simulation (Evrard et al.~\cite{evr+al96}),
\begin{equation}
\label{vir2}
r_{500}(T_{\rm X})=r_{10}(500, \Omega_{\rm M0})\left(
\frac{T_{\rm X}}{\rm{10KeV}}\frac{H_0^2}{H^2(z)}\right)^{1/2}\frac{1}{h},
\end{equation}
where $h$ is $H_0$ in units of 100 km s$^{-1}$Mpc$^{-1}$. The normalization $r_{10}$ in
Eq.~(\ref{vir2}) is the average radial scale of $10$ KeV clusters at density contrast
$\delta_{\rm c}=500$. Since the scaling law reflects the virial equilibrium within
$\delta_{\rm c} \sim {\cal{O}}(10^3)$, $r_{10}(500)$ is nearly insensitive to the peculiar
cosmology. We put $r_{10}(500,
\Omega_{\rm M0})= 1.24 \pm 0.08$ Mpc (Evrard et al.~\cite{evr+al96}).

The gas fraction at $r_{500}$, once known at a fixed physical radius $r_{\rm X}$, can be
evaluated using a mild, power law extrapolation of the data quoted at $r_{\rm X}$
(Evrard~\cite{evr97}),
\begin{equation}
\label{vir3}
f_{\rm ICM}[r_{500}(T_{\rm X})]=f_{\rm ICM}(r_{\rm X})\left[
\frac{r_{500}(T_{\rm X})}{r_{\rm X}}\right]^{\eta},
\end{equation}
with $\eta=0.17$ as derived from numerical simulations (Evrard et al.~\cite{evr+al96}). The
small value of $\eta$ also reduces the error on the gas fraction which propagates from an
uncertain determination of the normalization factor in Eq.~(\ref{vir2}). Using a different
normalization based on observations of the relatively relaxed cluster A1795, Mohr et
al.~(\cite{moh+al99}) found $r_{10}(500)
\approx 1.19$. Since $\eta = 0.17$, the relative variation on
$f_{\rm ICM}(500)$ between the two normalization is less then $1\%$.

Present X-ray observations are mostly sensitive to the inner cluster regions, but, future SZE
data would probe more exterior region, allowing an analysis on larger scales. There are some
evidences that the gas fraction increases by about 15\% from $r_{500}$ to $r_{200}$ (Ettori
\& Fabian \cite{ett+fab99}). Since the currently derived gas fraction profile increases
regularly from the inner part to the outer part, up to the virial radius and beyond (Sadat \&
Blanchard~\cite{sad+bla01}), the $\eta$ value at $\delta_{\rm c} \ll 500$ accordingly changes
but, whenever the ratio of $r_{\delta_{\rm c}}$ and $r_{\rm X}$ is within few percent of each
other, the effect of $\eta$ on the final result is really negligible (Cooray~\cite{coo98}).

The variations in the gas fraction within the virial regions are then quite small. At a given
temperature, the scatter in $f_{\rm ICM}$ is less than $20\%$, including both intrinsic
variations and measurements errors (Arnaud \& Evrard~\cite{arn+evr99};
Vikhlinin~\cite{vik+al99}).

The estimated gas mass fraction out to $r_{500}$ depends on cosmology through the angular
diameter distance, see Eq.~(\ref{frac6}), and the time dependent Hubble constant, see
Eq.~(\ref{vir2}). From Eqs.~(\ref{frac6},~\ref{vir2},~\ref{vir3}), we get
\begin{equation}
\label{vir4}
f_{\rm ICM}[r_{500}] \propto d_{\rm A}^{\beta -\eta} H(z)^{-\eta}.
\end{equation}
$f_{\rm ICM}[r_{500}]$ is a decreasing function of both $\Omega_{\rm
M0}$ and $H_0$. So, an increment in $\eta$ determines an overestimate
of these cosmological parameters.

Under the assumption that the gas fraction is time independent and constant for all clusters,
it is possible to constrain the cosmological parameters (Sasaki~\cite{sas96};
Pen~\cite{pen97}; Cooray~\cite{coo98}; Danos \& Pen~\cite{dan+pen98}; Rines et
al.~\cite{rin+al99}). The evolution of the gas mass fraction is still under discussion
(Schindler~\cite{schi99}; Ettori \& Fabian~\cite{ett+fab99}; Roussel et al.~\cite{rou+al00}).
An observed variation of $f_{\rm ICM}$ with the redshift would be explained by a wrong
assumption for the angular diameter distance and the cosmological parameters and/or a true
time evolution. In the near future, the study of the power spectrum of the secondary CMBR
anisotropies due to the thermal SZE by clusters of galaxies should give a discriminatory
signature of any possible evolution of the ICM mass fraction (Majumdar~\cite{maju01});
however, at the moment, numerical simulations do not suggest any evolution.

\section{Data sample}

The observational sample we consider is based on the X-ray data set in Mohr et
al.~(\cite{moh+al99}, hereafter MME) and Ettori \& Fabian~ (\cite{ett+fab99}; EF), on the SZE
data from Grego et al. (\cite{gre+al01}; GCR), and on the joint analysis of interferometric
SZE observations with X-ray imaging observations in Reese et al. (\cite{ree+al02}; RCJ). All
the published ICM mass fractions are given out to an angular radius within which the
signal-to-noise ratio is good enough and the problems deriving from extrapolation procedure
are really negligible (Sadat \& Blanchard~\cite{sad+bla01}).

In order to work with a homogeneous sample, we impose a cut on the
temperature, requiring $T_{\rm X} >5$~KeV. Cluster evolution is not an
entirely self similar process driven exclusively by gravitational
instability (David et al.~\cite{dav+al95}; Ponman et
al.~\cite{pon+al96}; MME). While massive clusters are less affected by
processes like galaxy feedback, low mass clusters may have lost gas as
a result of preheating and post-collapse energy input, so enhancing
ICM depletion within the virial region. The net result is that the ICM
mass fraction shows a mild increasing trend with the temperature. Both
observations (MME) and numerical simulations of the effect of
preheating (Bialek et al.~\cite{bia+al01}) show that $f_{\rm B}$ is
depressed below the cosmic mean baryon fraction in clusters with
$T_{\rm X}
\stackrel{<}{\sim} 3$~KeV. However, the trend between $f_{\rm B}$ and $T_{\rm X}$
is not clear (Roussel et al.~\cite{rou+al00}) and an analysis in
Arnaud \& Evrard~(\cite{arn+evr99}) of selected clusters with weak
cooling flow does not give clear statistical significance. Sanderson
et al~(\cite{san+al03}) found departures from a self-similar behaviour
in the scaling properties of a large sample of virialized haloes. Both
the relation between the gas density slope parameter and the
temperature and the gas fraction data reveal a flattening of the gas
density profiles in small sized haloes, consistent with energy
injection into the ICM by non-gravitational means. A clear trend for
cooler system to have a smaller gas fraction emerged, although, above
4~KeV, the significance of this correlation is weakened.

To be more conservative with respect to these shortcomings, we consider only clusters with
$T_{\rm X}
>5$~KeV, which are less affected by feedback from galaxy
formation and where $f_{\rm ICM}$ appears to be constant without opposite claims. This
criterion is passed by $28$ clusters from the sample in MME, $35$ from the sample in EF, $18$
from both the sample in GRC and RCJ

\section{Data analysis}

Median statistics provide a powerful tool to experimental data analysis. Few assumptions
about the data and their errors are required. Usual $\chi^2$ statistics assume that {\it i)}
experimental data are statistically independent; {\it ii)} there are no systematic effects;
{\it iii)} experimental errors are Gaussianly distributed; {\it iv)} the standard deviation
of these errors is known.  On the other hand, median statistics assume only hypotheses {\it
i)} and {\it ii)}.

To compute the likelihood of a particular set of cosmological parameters, we count how many
data points are above or below each cosmological model prediction and compute the binomial
likelihoods. Given a binomial distribution, if we perform $N$ measurements, the probability
of obtaining $k$ of them above the median is given by
\begin{equation}
\label{med1}
P(k)=\frac{2^{-N}N!}{k!(N-k)!}.
\end{equation}
We perform such a test on the baryonic gas fraction measured with X-ray data from MME and EF,
on the gas mass fraction from SZE observations from GCR and, finally, on the cosmological
distances measurements in RCJ

\subsection{Baryonic mass fractions}

We count the number of baryon mass fractions that are too heavy with respect to  (i.e.
greater than) the ratio $\Omega_{\rm B0} / \Omega_{\rm M0}$. ICM is not the only contribution
to the cluster baryon budget. Other contributions come from the luminous mass in galaxies,
intergalactic stars and a hypothetical baryonic dark matter. While the latter two terms can
be neglected (Ettori et al.~\cite{etto01}), the typical stellar contribution to the baryonic
mass is between $5$ and $20\%$ (White et al.~\cite{whi+al93}; Fukugita et
al.~\cite{fuk+al98}; Roussel et al.~\cite{rou+al00}). The ICM is the most extended mass
component in clusters, while the galaxies are the most centrally concentrated one (David et
al.~\cite{dav+al95}). There are some indications of a trend of increasing gas mass fraction
and decreasing mass in optically luminous matter with increasing mass of the clusters (David
et al.~\cite{dav+al95}; David~\cite{dav+al95}), but this claim is still debated (Roussel et
al.~\cite{rou+al00}). Lin et al.~(\cite{lin+al03}) found that the total baryon fraction is an
increasing function of the cluster mass. Estimating the stellar mass using $K$-band
luminosity, they also showed how the ICM to stellar mass ratio nearly doubles from low- to
high-mass clusters. However, our choice to consider only massive clusters with $T_{\rm X}
>5$~KeV gives some cautions in considering the total mass in galaxies as a fixed fraction of
the cluster gas. We have adopted the estimate in Fukugita et al.~(\cite{fuk+al98}), $f_{\rm
B}=(1+\left[0.18^{+0.10}_{-0.08}\right]h^{1/2})f_{\rm ICM}$.

A determination for $\Omega_{\rm B0}$ is required. One of the most
precise determination of the physical baryonic density is derived in
O'Meara et al.~(\cite{ome+al01}). By combining measurements of the
primeval abundance of deuterium, as inferred from high-redshift
Ly$\alpha$ systems, and theoretical predictions of the big-bang
nucleosynthesis, they found $\Omega_{\rm B0} h^2= 0.0205 {\pm} 0.0018$.
Measurements of the angular power spectrum of the CMBR also provide an
estimate of the baryon abundance. For the recent WMAP data (Spergel et
al.~\cite{spe+al03}), $\Omega_{\rm B0} h^2= 0.024 {\pm} 0.001$, depending
primarily on the ratio of the first to second peak heights. When
combined with other finer CMBR experiments and 2dFGRS measurements,
WMAP data gives $\Omega_{\rm B0} h^2= 0.022 {\pm} 0.001$ (Spergel et
al.~\cite{spe+al03}), in remarkable agreement with the result from
Ly$\alpha$ forest data. CMBR predictions depend on a multi-dimensional
fitting procedure involving all the cosmological parameters together,
included the parameters we want to determine, i.e. $H_0$ and
$\Omega_{\rm M0}$, so we prefer to adopt the estimate from O'Meara et
al.~(\cite{ome+al01}).

The ICM mass fraction in Eq.~(\ref{vir4}) is a decreasing function of
both $H_0$ and $\Omega_{\rm M0}$, so, increasing the observed value of
the baryon abundance, i.e. $\Omega_{\rm B0}$, determines an
underestimate of both $H_0$ and $\Omega_{\rm M0}$.

We use the ICM mass fractions to a fixed metric radius as reported in
Table~4 in MME, Table~ 2 in EF and Table~4 in GCR. Then, we
extrapolate to the virial radius assuming the reference cosmological
model ($\Omega_{\rm M0}=1$ and $h=0.5$ in MME and in EF and
$\Omega_{\rm M0}=0.3$ and $h=1$ in GCR); the prediction of a generic
cosmological model is obtained according to Eq.~(\ref{vir4}). We join
the X-ray derived mass fraction lists in MME and EF in a single data
sample with 63 data points. The 1-$\sigma$ and region 2-$\sigma$ in
the $\Omega_{\rm M0}$-$h$ plane are located by cosmological pairs
whose predictions are between 28 and 35 times ($68.65\%$\footnote{This
and the following probabilities are determined by adding the single
binomial likelihoods of $k$ overestimates in a data sample of $N$
entries, according to Eq.~(\ref{med1}). The 1-$\sigma$ and the
2-$\sigma$ confidence region are chosen to be symmetric around the
median and such that the total probability inside them is just larger
than 68.3\% and 95.4\%, respectively.} confidence region) or between
24 and 39 times ($95.70\%$), respectively, above $\Omega_{\rm
B0}/\Omega_{\rm M0}$.

\begin{figure}
   \resizebox{\hsize}{!}{\includegraphics{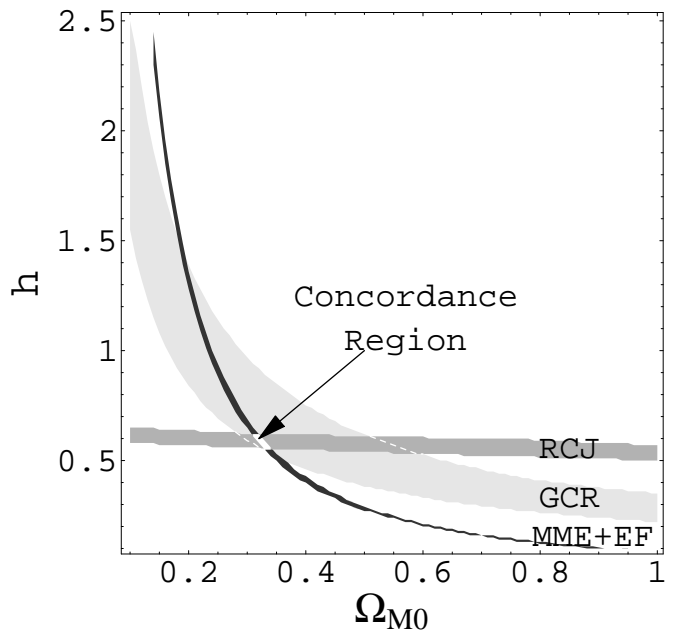}}
   \caption{Concordance region (white) at the 1-$\sigma$ level. Different
   gray-scaled patches show constraints from different data samples. Regions
   labelled with RCJ, GRC and MME+EF denote constraints from cosmological distances,
   baryonic mass fractions from SZE measurements and baryonic mass fractions from X-ray data, respectively.}
    \label{conc1}
\end{figure}

The SZE derived mass fraction data sample provides 18 entries. Now, the 1-$\sigma$ and
2-$\sigma$ regions in the $\Omega_{\rm M0}$-$h$ plane are located by points with a number of
overestimates between 7 and 11 ($76.2\%$) and 5 and 13 ($96.9\%$), respectively. Since the
different dependence on the cosmological parameters, the results from X-ray and SZE
measurements locate independent regions in the $\Omega_{\rm M0}$-$h$ plane, see
Fig.~(\ref{conc1}). In any case the points in the SZE sample are not enough to significantly
constrain cosmological parameters from estimates of gas fraction alone.

\begin{figure}
   \resizebox{\hsize}{!}{\includegraphics{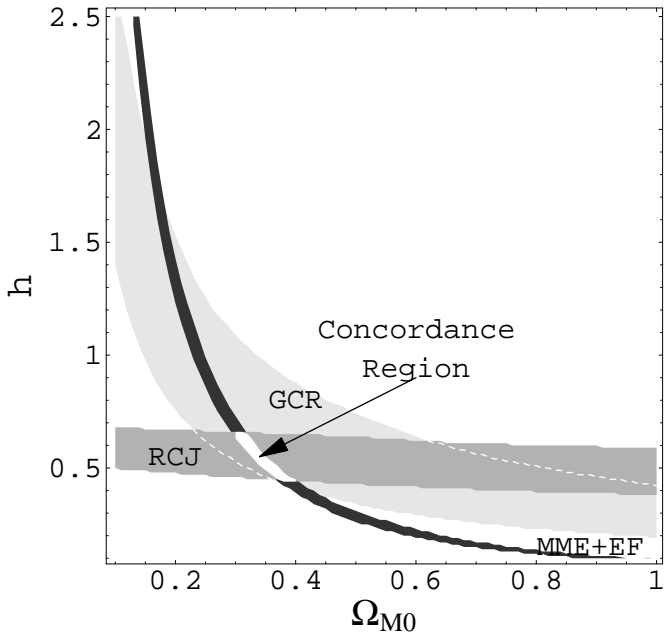}}
   \caption{Concordance region (white) at the 2-$\sigma$ level. Different gray-scaled
   patches show constraints from different data samples. Regions
   labelled with RCJ, GRC and MME+EF denote constraints from cosmological distances,
   baryonic mass fractions from SZE measurements and baryonic mass fractions from X-ray data, respectively.}
    \label{conc2}
\end{figure}

\subsection{Cosmological distances}
As seen in Section~\ref{sec2}, when combining X-ray and SZE data, a third constrain can be
obtained without referring to any value of $\Omega_{\rm B0}$. Reese et al.~(\cite{ree+al02})
determined the distances to 18 galaxy clusters with redshift ranging from $z \sim 0.14$ to
$0.78$. Now, for each cosmological model, we count the number of clusters that are too
distant, i.e the number of cluster whose distance calculated  according to Eq.~(\ref{dist1})
for a set of cosmological parameters is greater than the measured value. The confidence
region in the $\Omega_{\rm M0}$-$h$ plane are determined with the same criteria as for the
SZE derived mass fraction. Despite the numbers of entries for the cosmological distances and
for the baryonic mass fraction from SZE data are the same, the former does better since the
difference dependence on cosmological parameters. The test on the cosmological distances,
very sensitive to the Hubble constant, provides a nearly orthogonal constraint to the method
of the baryonic fraction. At the 1-$\sigma$ level, we get $0.50
\stackrel{<}{\sim} h \stackrel{<}{\sim} 0.66$; the 2-$\sigma$ confidence range is
$0.38 \stackrel{<}{\sim} h \stackrel{<}{\sim} 0.70$. Using a $\chi^2$
statistics, where the statistical uncertainties have been obtained
combining in quadrature and then averaging not Gaussian, asymmetric
errors, Reese et al.~(\cite{ree+al02}) found, as $68.3\%$ confidence
range, $0.51
\stackrel{<}{\sim} h \stackrel{<}{\sim} 0.64$, nearly as constraining as our result.

\subsection{Concordance analysis}

To evaluate the allowed range of $\Omega_{\rm M0}$ and $h$, we have to
consider together the three independent constraints. We apply a
concordance analysis, only retaining the pairs of cosmological
parameters which lie within 1-$\sigma$ or 2-$\sigma$ of each
individual constraint (Wang et al.~\cite{wan+al00}). This procedure is
conservative with respect to possibly not well controlled systematic
errors and puts in evidence the most effective constraints in
delimiting the allowed range. As can be seen from
Figs.~(\ref{conc1},~\ref{conc2}), constraints from cosmological
baryonic fraction are nearly orthogonal to that from cosmological
distances. At the 1-$\sigma$ level, Fig.~(\ref{conc1}), we get $0.31
\stackrel{<}{\sim} \Omega_{\rm M0}
\stackrel{<}{\sim} 0.34$ and $0.55 \stackrel{<}{\sim} h
\stackrel{<}{\sim} 0.62$. At the 2-$\sigma$ level,
Fig.~(\ref{conc2}), we get $0.30 \stackrel{<}{\sim} \Omega_{\rm M0}
\stackrel{<}{\sim} 0.40$ and $0.44 \stackrel{<}{\sim} h
\stackrel{<}{\sim} 0.66$.

\section{Systematic effects}

Several systematic uncertainties are involved when deriving cosmological parameters from the
cluster gas mass observations.

\subsection{ICM depletion}
\label{deple}
Shocks during cluster formation can drive an ICM depletion within $r_{500}$. Numerical
simulations (Frenk et al.~ \cite{fre+al99}) and observations (Sanderson et
al.~\cite{san+al03}) show that the gas distribution is more extended than the dark matter
one. The extended gas structure implies a weakly rising baryon fraction with radius
(Evrard~\cite{evr97}; EF), $f_{\rm ICM}\sim r^{\eta}$, see Eq.~(\ref{vir3}), and this
increasing trend is stronger in less massive clusters (Schindler~\cite{schi99}). The ratio
$\Upsilon$ of the enclosed baryon fraction to the universal cosmic value within a fixed
density contrast,
\begin{equation}
\label{depl1}
\Upsilon(\delta_{\rm c})\equiv \left( \frac{\Omega_{\rm B0}}{\Omega_{\rm M0}}\frac{1}{f_{\rm B}(\delta_{\rm c})}
\right)^{-1},
\end{equation}
has been calibrated by simulations. A modest overall baryonic diminution, about $10\%$, has
been found (Evrard~\cite{evr97}; Frenk et al.~\cite{fre+al99}).

\subsection{Density clumping}
\label{clump}
Small-scale density fluctuations on X-ray measurements of the ICM mass arising from accretion
events and major mergers can introduce a bias in the determination of $f_{\rm ICM}^X$.
Density clumping on a scale less than the resolution of the images causes an enhancement of
the X-ray brightness by a factor $C_{\rm n}=\langle n_{\rm e}^2\rangle/\langle n_{\rm
e}\rangle^2$ with respect to a uniform smooth atmosphere. Since the surface brightness
profile is proportional to the square of the central gas density, see Eq.~(\ref{frac3}),
$f_{\rm ICM}^X$ is overestimated by $C_{\rm n}^{1/2}$. Numerical hydro-simulations show that
$C_{\rm n}^{1/2} = 1.16
\pm 0.01$ (Mathiesen et al.~\cite{mat+al99}). However, the spread in
this value, on a cluster-by-cluster basis, can be much larger than this.

Since the value of $\langle n_{\rm e} \rangle$ is not changed by density clumping, the
estimate of $f_{\rm ICM}$ based on the SZE is not affected by this systematic effect.

Currently, there is no observational evidence of significant clumping in galaxy clusters
(RCJ).

\subsection{Other effects}
Several other systematic effects can affect the gas mass fraction measurements. A magnetic
field in the cluster plasma might support a non thermal component in the X-ray emission. It
can be effective in the core regions of the cluster but it is unlikely that it plays a major
role out to the hydrostatic radius (Cooray~\cite{coo98}). The non-isothermality of the ICM
has also been taken into account (Ettori et al.~\cite{etto01}; Puy et al.~\cite{puy+al00}).
The presence of a temperature gradient increases the gas fraction estimated at the virial
radius $r_{500}$, whereas correcting by the contamination of a magnetic field reduces $f_{\rm
ICM}$ (Ettori et al.~\cite{etto01}).

Heating and cooling process may also act in galaxy clusters, but they do not change in a
significant way the estimate of the mean $f_{\rm ICM}$ in a numerous enough cluster sample
(Cooray~\cite{coo98}). Cooling processes can alter the ICM profiles. Since the SZE depends
essentially on the pressure profile, a cooling flow can lead to an underestimation of the
cosmological distance (Majumdar \& Nath~\cite{ma+na00}). Even after excluding $\sim 80\%$ of
the cooling-flow region from the analysis, a $\sim 10\%$ overestimation of $H_0$ may be in
order.

Cluster geometry introduces an important uncertainty in SZE- and
X-ray-derived quantities (Cooray~\cite{coor98b}; Puy et
al.~\cite{puy+al00}). Complicated cluster structure and projection
effects cannot currently be disentangled. Projection effects of
aspherical cluster modelled with a spherical geometry broaden the
distribution of measured gas mass fraction and should be corrected by
taking into account the distribution of ellipticities for the cluster
sample (Cooray~\cite{coor98b}). The effects of asphericity contribute
significantly to the distance uncertainty for each cluster, but the
determination of the Hubble constant from a large sample of clusters
is not believed to be significantly biased (RCJ).

These various effects do not act all in the same direction; furthermore, they require a very
detailed modelling in order to correct for a not very significant amount. Since an ensemble
of hydro-dynamical cluster simulations has shown that the departures from an isothermal,
spherical ICM do not introduce serious errors (GRC), we have not considered them in our
analysis.

Considering the effects discussed in Section~(\ref{deple},~\ref{clump}), ICM mass fraction
should be corrected according to
\begin{equation}
\label{corr1}
f_{\rm B}^{\rm X}=f_{\rm ICM}^X\left[ 1+\frac{f_{\rm gal}}{f_{\rm
ICM}}\right]\frac{1}{\Upsilon (500)}\frac{1}{C_{\rm n}^{1/2}},
\end{equation}
and
\begin{equation}
\label{corr2}
f_{\rm B}^{\rm SZ}=f_{\rm ICM}^{\rm SZ}\left[ 1+\frac{f_{\rm gal}}{f_{\rm
ICM}}\right]\frac{1}{\Upsilon (500)}.
\end{equation}

\section{Conclusions}
The simple argument presented here, based on the different dependence of the galaxy cluster
ICM mass estimates on the cosmological parameters as derived either from SZE measurements or
from X-ray observations, has made it possible to infer both the value of the pressureless
matter density parameter and the Hubble constant. The only additional information is the
value of the universal baryonic matter density. To our knowledge, this is the first time that
gas mass estimates are used to derive $\Omega_{\rm M0}$ and $H_0$ together.

We have followed median statistics. This technique assumes only that the measurements are
independent and free of systematic errors. Physical quantities in literature, such as X-ray
emission temperature ICM mass fractions and cosmological distances, are often presented with
asymmetric errors bars. Supposing these errors following a Gaussian distribution with known
standard deviation, as required by usual $\chi^2$ analysis, is a hard hypothesis. On the
other hand, median statistics, based on fewer assumptions, can provide very interesting and
constraining results, as shown in several astrophysical cases (Gott et al.~\cite{got+al01};
Avelino et al.~\cite{ave+al02}; Chen \& Ratra~\cite{ch+ra03}). To minimize the role of
systematic uncertainties, we have performed a concordance analysis of the data (Wang et
al.~\cite{wan+al00}).

Our estimates agree with the currently favoured model of Universe
(Wang et al.~\cite{wan+al00}), derived from observational constraints
such as measurements of the anisotropy of the CMBR (Spergel et
al.~\cite{spe+al03}), large-scale structure observations (Peacock et
al.~\cite{pe&al01})and evidences coming from type Ia supernovae (Riess
et al. \cite{ri&al98}; Perlmutter et al. \cite{pe&al99}).

We have found a value of $\Omega_{\rm M0}$ in full agreement with the bulk of evidences
(Harun-or-Rashid \& Roos~\cite{ha+ro01}; Chen \& Ratra~\cite{ch+ra03}). Median statistics
analyses of various collections of measurements has been used in Chen \&
Ratra~(\cite{ch+ra03}) to determine an estimate of $\Omega_{\rm M0}$. They found $0.20
\stackrel{<}{\sim} \Omega_{\rm M0}
\stackrel{<}{\sim} 0.35$ at two standard deviations, in full agreement
with our estimate $0.30 \stackrel{<}{\sim} \Omega_{\rm M0}
\stackrel{<}{\sim} 0.40$.

Our determination of the Hubble constant is independent of the local extragalactic distance
scale. We find $0.44 \stackrel{<}{\sim} h
\stackrel{<}{\sim} 0.66$ at 2-$\sigma$, in agreement with the estimate of the Hubble Space Telescope Key Project, $h
=0.72 {\pm}8$ (Freedman et al.~\cite{fre+al01}).
Together with SZE-derived distances, time delays produced by lensing of quasars by foreground
galaxies also provide a tool to determine $H_0$ independent of the extragalactic distance
ladder. This technique suffers by very uncertain systematics but gives results in agreement
with our result (Witt et al.~\cite{wit+al00}). We remark how both these two global methods
tend to yield smaller estimates of $H_0$ than the determination by CMBR measurements: WMAP
data give $h=0.72 \pm 0.05$ (Spergel et al.~\cite{spe+al03}). The value of the Hubble
constant, as derived applying median statistics to a collection of nearly all available
pre-mid-1999 estimates, is $h=0.67 {\pm}0.02$ (95\% statistical) ${\pm}0.05$ (95\% systematic) (Gott
et al.~\cite{got+al01}), in agreement with our result.

In the near future, by increasing the SZE data sample, the estimates of the cosmological
parameters obtained with our method should greatly improve.

\end{document}